# Automatic Generation Control Considering Uncertainties of the Key Parameters in the Frequency Response Model

Likai Liu, *Student Member, IEEE,* Zechun Hu, *Senior Member, IEEE,* and Asad Mujeeb, *Student Member, IEEE*

*Abstract*—The highly fluctuated renewable generations and electric vehicles have undergone tremendous growth in recent years. The majority of them are connected to the grid via power electronic devices, resulting in wide variation ranges for several key parameters in the frequency response model (FRM) such as system inertia and load damping factor. In this paper, an automatic generation control (AGC) method considering the uncertainties of these key parameters is proposed. First, the historical power system operation data following large power disturbances are used to identify the FRM key parameters offline. Second, the offline identification results and the normal operation data prior to the occurrence of the disturbance are used to train the online probability estimation model of the FRM key parameters. Third, the online estimation results of the FRM key parameters are used as the input, and the model predictive-based AGC signal optimization method is developed based on distributionally robust optimization (DRO) technology. Case studies conducted on the IEEE 118-Bus System show that the proposed AGC method outperforms the widely utilized PI-based control method in terms of performance and efficiency.

## Nomenclature

| | |
|---|---|
| $H$ | Power system inertia factor. |
| $D$ | Load damping factor. |
| $R_n$ | Droop coefficient of regulation resource $n$. |
| $\alpha_n$ | Participation factor of regulation resource $n$. |
| $B$ | Frequency bias factor of the system. |
| $K(s)$ | Transfer function of the AGC controller. |
| $T_n^{\mathrm{G}}$ | Governor time constant of steam turbine $n$. |
| $T_n^{\mathrm{C}}$ | Turbine time constant of steam turbine $n$. |
| $T_n^{\mathrm{R}}$ | Reheater time constant of reheat steam turbine $n$. |
| $F_n^{\mathrm{H}}$ | High-pressure turbine fraction of reheat steam turbine $n$. |
| $T_n^{\mathrm{V}}$ | Converter time constant of BESS $n$. |
| $\Delta f$ | System frequency deviation. |
| $\Delta P_n^{\mathrm{M}}$ | Change of the mechanical power of steam turbine $n$. |
| $\Delta P_n^{\mathrm{C}}$ | Change of the steam intake of steam turbine $n$. |
| $\Delta P_n^{\mathrm{G}}$ | Change of the value opening of steam turbine $n$. |
| $\Delta P_n^{\mathrm{E}}$ | Change of the electric power of BESS $n$. |
| $\Delta P^{\mathrm{R}}$ | Total secondary frequency control signal. |
| $\Delta P^{\mathrm{L}}$ | Load power disturbance. |
| $\Delta P^{\mathrm{NL}}$ | Netload power disturbance. |
| $\overline{\Delta f}, \underline{\Delta f}$ | Upper and lower bounds of the frequency deviation. |
| $\overline{P^R}, \underline{P^R}$ | Upper and lower bounds of the total secondary frequency regulation signal. |

## I. Introduction

POWER systems are currently witnessing unprecedented changes. Environmental and sustainability concerns have resulted in the replacement of a sizable portion of conventional fossil fuel-based power plants with renewable energy resources, which may eventually become the primary power source in the future under the carbon neutrality plan. Nonetheless, the high penetration of renewable generations also brings challenges to the power system frequency control.

Automatic generation control (AGC) is an essential part of the power system frequency control. Although the PI-based AGC method has been widely used in modern power systems for its simplicity and stability [1], the lack of consideration for changeable control condition impairs its control performance. Several other control theories, such as fuzzy logic control [2], robust control [3], and sliding mode control [4], have also been applied in the controller design to improve the AGC performance. The high penetration of renewable energy lowers frequency control performance and jeopardizes frequency security, but the above-mentioned methods cannot explicitly handle the frequency related constraints. Thus, they may face challenges with steadily increasing renewable generations [5].

Model predictive control (MPC) can achieve effective frequency regulation with the control constraints ensured, which has been used in AGC [6]–[10]. The efficiency and performance of MPC are highly relay on accurate system modeling. In the frequency response model (FRM), some key parameters are time-varying and have significant effects on frequency control [11], such as system inertia and load damping factor. However, the aforementioned model predictive-based AGC methods are developed without proper regard for the uncertainties associated with these parameters.

The FRM key parameters have shown larger ranges of variation after the large-scale integration of converter-interfaced resources [12], [13], such as the centralized wind/solar power plants, behind-the-meter distributed power generations, and electric vehicles. Unlike traditional generation and load resources, most of these converter-based resources currently do

The authors are with the Department of Electrical Engineering, Tsinghua University, Beijing, 100084, China (e-mail: zechhu@tsinghua.edu.cn).

not provide inertial support and frequency regulation service for the power system [14]. The increasing penetrations of these resources have led to significant heterogeneities in power system generation and load resources. Moreover, their power injections/consumptions have strong randomness and volatility, which have brought large uncertainties into the FRM key parameters. Given the non-negligible influences of time-varying FRM key parameters on the AGC, identifying these parameters during online operation becomes essential for implementing MPC technology.

The existing literature on power system parameter identification mainly centers on system inertia [11], and it can be further divided into offline and online identification methods. The offline identification method is typically based on the historical operation data after large power disturbances, such as generator outages [15], [16], which can help to reduce the influences from the dead-band of the governor and noise in the PMU data. However, offline identification cannot deliver the real-time identification result.

Online identification can be further divided into two kinds of methods. The first kind is to calculate the inertia contributed by all online traditional generators (TGs) [17], [18], which only yields a lower bound of the system inertia. The second method is to use the power disturbance data recorded by the PMU to identify system inertia during the online stage [19], [20]. Its accuracy is questionable because the large power disturbance is quite rare, and the online parameter identifications are base on small power disturbances in most instances.

To compensate for the shortcomings of previous studies, this paper proposes an model predictive-based AGC method which considers the uncertainties of the FRM key parameters. First, a systematic parameter identification method is proposed to online estimate the probability distributions of the key parameters. On this basis, the distributionally robust optimization (DRO) theory is utilized to develop the AGC signal optimization method to handle the potential inaccuracies in the estimated probability distributions. The major contributions of this work are summarized as follows.

1) A systematic FRM key parameter identification method is proposed in this research, which organically combines the offline and online parameter identification methods while retaining their respective advantages.
2) A new framework is proposed that combines the probability estimation and DRO methods, with the DRO's ambiguity set established based on the probability estimation result.
3) Based on the MPC and DRO theories, an AGC signal optimization model considering the uncertainties of the FRM key parameters is built and transformed into a quadratically constrained quadratic programming problem which can be solved efficiently.

The remainder of this paper is organized as follows. Section II introduces the power system FRM. Section III presents the offline and online FRM key parameter identification methods. The model predictive-based AGC signal optimization method is developed in Section IV. Numerical experiments are given in Section V. Section VI concludes this paper.

## II. FREQUENCY RESPONSE MODEL OF POWER SYSTEM

### A. State Space Equation

In this paper, we study a power system with a single balancing area, whose FRM including both primary and secondary frequency controls is depicted in Fig. 1. The frequency regulation resources in this system include $N^{\mathrm{rh}}$ reheat steam turbines, $N^{\mathrm{nr}}$ non-reheat steam turbines, and $N^{\mathrm{es}}$ battery energy storage systems (BESSs). These regulation resources are arranged according to their types.

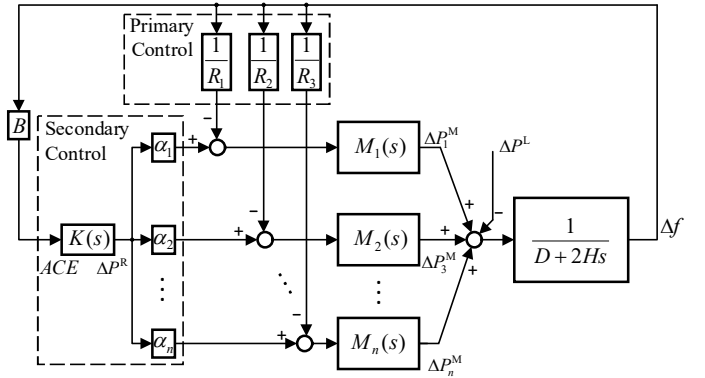

Fig. 1. Block diagram of an isolated balancing area containing primary and secondary frequency controls.

The block $M_n(s)$ in Fig. 1 represents the transfer function of the regulation resource, and the detailed models for different regulation resources are shown in Fig. 2 (a)-(c).

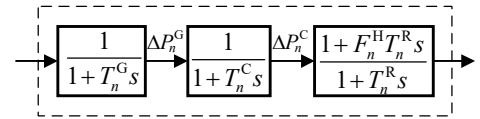

(a) Frequency response model for a reheat steam turbine.

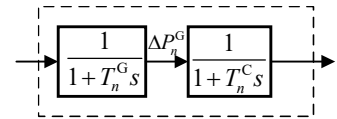

(b) Frequency response model for a non-reheat steam turbine.

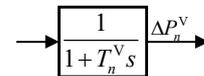

(c) Frequency response model for a battery energy storage system.

Fig. 2. Frequency response models for various frequency regulation resources.

Based on the models shown in Fig. 1 and Fig. 2, the state space model of the frequency control system can be formulated as follows:

$$\dot{x} = \mathcal{A}x + \mathcal{B}u, \qquad (1)$$

$x = [\Delta f, \Delta P_1^{\mathrm{M}}, \Delta P_1^{\mathrm{C}}, \Delta P_1^{\mathrm{G}}, \cdots, \Delta P_{1+N^{\mathrm{rh}}}^{\mathrm{M}}, \Delta P_{1+N^{\mathrm{rh}}}^{\mathrm{G}}, \cdots,$
$\Delta P_1^{\mathrm{E}}, \cdots, \Delta P_{N^{\mathrm{es}}}^{\mathrm{E}}]^{\top}$, $u = [\Delta P^{\mathrm{R}}, \Delta P^{\mathrm{NL}}]^{\top}$.

The matrices $\mathcal{A}$ and $\mathcal{B}$ are given in Appendix A.

## B. Time-Domain System Frequency Response Model

After establishing the system FRM, the Laplace transform is used to obtain the time-domain expression of the state variables. First, take the Laplace change on both sides of the state space equation (1), as

$$sX(s) - x(0) = \mathcal{A}X(s) + \mathcal{B}U(s). \quad (2)$$

where $x(0)$ is a vector composed of the initial values of the state variables. Secondly, the frequency domain expression of the state variables is derived as follows:

$$X(s) = (s\mathbf{I} - \mathcal{A})^{-1}x(0) + (s\mathbf{I} - \mathcal{A})^{-1}\mathcal{B}U(s). \quad (3)$$

where $\mathbf{I}$ is the unit matrix with the proper dimension. Finally, the inverse Laplace transform is used to calculate the time-domain expression of the state variables, as

$$\begin{aligned}x(t) &= \mathcal{L}^{-1}\left[(s\mathbf{I}-\mathbf{A})^{-1}x(0)\right] + \mathcal{L}^{-1}\left[(s\mathbf{I}-\mathcal{A})^{-1}\mathcal{B}U(s)\right] \\ &= e^{\mathcal{A}t}x(0) + \int_0^t e^{\mathcal{A}(t-\tau)}\mathcal{B}u(\tau)d\tau\end{aligned} \quad (4)$$

In this paper, the above formulation will be employed to develop the FRM key parameter identification method and model predictive-based AGC method.

## III. IDENTIFICATION OF FRM KEY PARAMETERS

In the FRM established in the previous section, the parameters $H$ and $D$ have notable influences on AGC, and hence *they are defined as the FRM key parameters* in this study. Both $H$ and $D$ are difficult to obtain even the inertia contributed by online TGs can be easily calculated, because some loads have inertia and damping effects, and the load composition is complex and changeable especially after the large-scale integration of converter-based devices. To guarantee the efficiency of the model predictive-based AGC method, these FRM key parameters should be identified during the power system online operation. This section proposes a systematic FRM key parameter identification method comprised of offline and online identifications.

### A. Offline Identifications of FRM Key Parameters

After a large power disturbance, the primary frequency control takes effect first, followed by the secondary frequency control after a certain period of time. Therefore, at the start of the disturbance, the effects of the secondary frequency control can be ignored, and the time-domain expression of the state variables is written as follows according to (4):

$$x^{\text{pri}}(t) = e^{\mathcal{A}t}x(0) + \int_0^t e^{\mathcal{A}(t-\tau)}\mathcal{B}[2]\Delta P^{\text{NL}}(\tau)d\tau \quad (5)$$

where $\mathcal{B}[2]$ represents the second column of matrix $\mathcal{B}$.

Replacing $t$ in (5) by the sampling period $\Gamma$ of the historical operation data, and assuming that the power disturbance between two adjacent sampling points changes linearly, then the state variables after one sampling period are formulated as

$$\begin{aligned}x(\Gamma) = &e^{\mathcal{A}\Gamma}x(0) + \Delta P^{\text{NL}}(0)\int_0^\Gamma e^{\mathcal{A}(\Gamma-\tau)}\mathcal{B}[2]d\tau \\ &+ \frac{\Delta P^{\text{NL}}(\Gamma) - \Delta P^{\text{NL}}(0)}{\Gamma}\int_0^\Gamma e^{\mathcal{A}(\Gamma-\tau)}\mathcal{B}[2]\tau d\tau\end{aligned} \quad (6)$$

On this basis, the time-domain transform function of the state variables from the $k^{\text{th}}$ to the $(k+1)^{\text{th}}$ data sampling instants is obtained by replacing $x(0)$ and $x(\Gamma)$ with $x_k$ and $x_{k+1}$.

$$\begin{aligned}x_{k+1} = &e^{\mathcal{A}\Gamma}x_k + \Delta P_k^{\text{NL}}\int_0^\Gamma e^{\mathcal{A}(\Gamma-\tau)}\mathcal{B}[2]d\tau \\ &+ \frac{\Delta P_{k+1}^{\text{NL}} - \Delta P_k^{\text{NL}}}{\Gamma}\int_0^\Gamma e^{\mathcal{A}(\Gamma-\tau)}\mathcal{B}[2]\tau d\tau \\ = &\mathcal{A}^{\text{pri}}x_k + \mathcal{B}^{\text{pri}}u_k\end{aligned} \quad (7)$$

where matrices $\mathcal{A}^{\text{pri}}$, $\mathcal{B}^{\text{pri}}$, and $u_k$ are given in Appendix A.

Assuming that the system inertia and load damping factor do not change within the first minute following the occurrence of the power disturbance, then they can be simultaneously identified by using the least square method based on the power disturbance and system frequency data, as

$$\begin{aligned}\textbf{P1:} \quad &\min_{H_i,D_i} \sum_{k=1}^{K_i}\left(\Delta\hat{f}_i[k] - \Delta f_i[k]\right)^2 \\ &\text{s.t.:} \quad x_{i,k+1} = \mathcal{A}_i^{\text{pri}}x_{i,k} + \mathcal{B}_i^{\text{pri}}u_{i,k}, 0 \leq k < K_i\end{aligned} \quad (8)$$

where $k$ and $K_i$ denote the index and number of data points recorded in the $i^{th}$ power fluctuation event, $\Delta\hat{f}_i[k]$ and $\Delta f_i[k]$ are actual and calculated frequency deviations.

**P1** is a nonlinear programming problem, but fortunately, the dimension of its decision variables is low. As a result, the interior point method becomes a high-efficiency method for solving it, and a locally optimal solution can be certainly found. Furthermore, the quality of the locally optimal solution can be improved by presetting the search ranges of the FRM key parameters according to the operational conditions and experiences. For example, the lower bound of the system inertia can be set as the inertia contributed by all online TGs. The case study section tests the proposed offline parameter identification method and shows that it has good accuracy.

*Remarks*: As stated earlier, the power fluctuations used for the offline identification should have a significant amplitude. The generator tripping events are utilized to estimate the system inertia in [21] and [22], but the generator failure events are rare in practice. To increase the number of power disturbance events suitable for the parameter estimation, Schiffer *et al.* proposed to employ unit start and shutoff events for system inertia identification [23]. Renewable energy's total installed capacity has increased significantly in recent years. Renewable energy generation is extremely fluctuating, and its power ramp events can also be used to identify the FRM key parameters. Thus, the power sudden change events caused by TGs and renewable generations are utilized to offline estimate the key parameters in this study.

## B. Online Probability Estimations of FRM Key Parameters

This subsection develops a deep learning-based method to online estimate the probability distributions of the FRM key parameters. The probability estimation is adopted because it not only avoids the large error probably induced by the point estimation but also provides more information for the subsequent AGC signal optimization model. The normal operation data before the historical large power disturbances and the offline parameter identification results constitute the training samples, and they are used to train the online FRM key parameter estimation network. The advantage of this framework lies in utilizing the accurate offline identification results to train the online estimation network.

The large power fluctuation is rare in power system operation, so the online parameter identifications are typically based on small power disturbances. Even though the magnitude of the disturbance is relatively small, the power fluctuation and frequency data still contain information about the system inertia and load damping factor. Moreover, the load inertia and damping are directly related to the load power. Thus, the data of load power, renewable generations, and system frequency are used as the input of the online probability estimation model of the FRM key parameters. Given that TGs provide a sizeable share of system inertia, this portion of inertia is computed and used as an input of the estimation model. In each data sampling instant, the above information composes a vector $\bm{Y}$. The input of the online estimation model contains the operation data in the past $T^{\text{IN}}$ sampling instants.

The load power, renewable generations, and system frequency data are all time-series data. Long short-term memory (LSTM) is a time recurrent neural network that excels in dealing with the time-series data [24]. Therefore, a two-layer LSTM network is utilized to process the above data, as shown in Fig. 3. The output of the LSTM network is the characteristic variables $\{F_1, F_2, \cdots, F_U\}$, which are inputted into a simple one layer fully-connected neural network to estimate the distributions of the FRM key parameters. The final outputs of the estimation model are the quantiles of the FRM key parameters, which correspond to cumulative probabilities ranging from 0.5 % to 99.5 % with a 1% interval.

The Pinball loss function is commonly used as the loss function to train the probability estimation model [25]. The Pinball loss function of the quantile corresponding to the cumulative probability $q$ is given as an example:

$$L_q(x^* - x) = \begin{cases} q \cdot (x^* - x), & x^* \geq x, \\ (1-q) \cdot (x - x^*), & x^* < x, \end{cases} \quad (9)$$

where $x^*$ and $x$ are the actual value and predicted quantile of a FRM key parameter. The loss function of this probability estimation model is the sum of the loss function over each quantile, as

$$L = \sum_{q=0.5\%}^{99.5\%} [L_q(H^* - H_q) + L_q(D^* - D_q)]. \quad (10)$$

The probability estimation model of the FRM key parameters is trained by minimizing the above loss function based on gradient descent and back propagation algorithms [26].

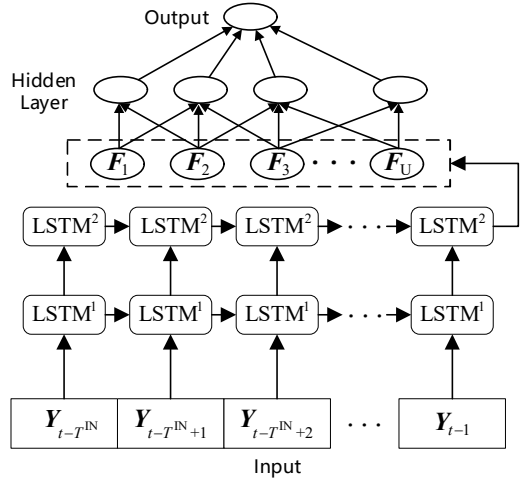

Fig. 3. Schematic of probability estimation model of the FRM key parameters.

## IV. MODEL PREDICTIVE-BASED AUTOMATIC GENERATION CONTROL

### A. Automatic Generation Control Model in Time-Domain

The AGC signal is viewed as a step signal in the time-domain system FRM established in Section II, and the power disturbance is simplified as a ramp signal, i.e., the disturbance changes linearly within one AGC period. Thus, the state variables in an AGC period are formulated as

$$\begin{aligned}\bm{x}(t) =& \mathrm{e}^{\bm{\mathcal{A}}t}\bm{x}(0) + \Delta P^{\text{R}} \int_0^t \mathrm{e}^{\bm{\mathcal{A}}(t-\tau)} \bm{\mathcal{B}}[1] \mathrm{d}\tau \\ &+ \frac{\Delta P^{\text{NL}}(t) - \Delta P^{\text{NL}}(0)}{t} \int_0^t \mathrm{e}^{\bm{\mathcal{A}}(t-\tau)} \bm{\mathcal{B}}[2] \tau \mathrm{d}\tau \\ &+ \Delta P^{\text{NL}}(0) \int_0^t \mathrm{e}^{\bm{\mathcal{A}}(t-\tau)} \bm{\mathcal{B}}[2] \mathrm{d}\tau, \end{aligned} \quad (11)$$

where $\bm{\mathcal{B}}[1]$ denotes the first column of matrix $\bm{\mathcal{B}}$.

By taking $t$ as the AGC period $\Upsilon$, as well as replacing $\bm{x}_z$ and $\bm{x}_{z+1}$ with $\bm{x}(0)$ and $\bm{x}(\Upsilon)$, the time-domain transform function of the state variables from the beginning to end of an AGC period is written as follows:

$$\begin{aligned}\bm{x}_{z+1} =& \mathrm{e}^{\bm{\mathcal{A}}\Upsilon}\bm{x}_z + \Delta P_z^{\text{R}} \int_0^{\Upsilon} \mathrm{e}^{\bm{\mathcal{A}}(\Upsilon-\tau)} \bm{\mathcal{B}}[1] \mathrm{d}\tau \\ &+ \frac{\Delta P_{z+1}^{\text{NL}} - \Delta P_z^{\text{NL}}}{\Upsilon} \int_0^{\Upsilon} \mathrm{e}^{\bm{\mathcal{A}}(\Upsilon-\tau)} \bm{\mathcal{B}}[2] \tau \mathrm{d}\tau \\ &+ \Delta P_z^{\text{NL}} \int_0^{\Upsilon} \mathrm{e}^{\bm{\mathcal{A}}(\Upsilon-\tau)} \bm{\mathcal{B}}[2] \mathrm{d}\tau \\ =& \bm{\mathcal{A}}^{\text{sec}} \bm{x}_z + \bm{\mathcal{B}}^{\text{sec}} \bm{u}_z \end{aligned} \quad (12)$$

where matrices $\bm{\mathcal{A}}^{\text{sec}}$, $\bm{\mathcal{B}}^{\text{sec}}$ and $\bm{u}_z$ are given in Appendix A.

### B. Construction of DRO Ambiguity Set

The actual probability distributions of the FRM key parameters may differ from the estimated distributions due to potential inaccuracies. This study employs the DRO method to construct the AGC signal optimization model to ensure that the proposed control method performs well under actual probability distributions.

There are mainly three types of DRO ambiguity sets: moment-based ambiguity set, metric-based ambiguity set, and scenario-based ambiguity set. In this study, the last kind of ambiguity set is adopted and combined with the probability estimation method: First, each quantile derived from the probability estimation is considered as a scenario, and the overall predicted quantiles comprise the empirical scenario set. Then, the ambiguity set is constructed by limiting the differences between the scenario weight vector $\boldsymbol{\omega}$ of a candidate scenario set and that of the empirical scenario set $\boldsymbol{\omega}^0$. Besides, the sum of the scenario weights should equal to 1, as

$$\widehat{\mathcal{P}} \triangleq \left\{ \boldsymbol{\omega} \,|\, \eta^{\min} \leq \boldsymbol{\omega} - \boldsymbol{\omega}^0 \leq \eta^{\max}, \boldsymbol{I}^\top \boldsymbol{\omega} = 1 \right\}. \quad (13)$$

where $\eta^{\max}$ and $\eta^{\min}$ are the upper and lower bounds of the scenario weight deviation.

The combination of probability estimation method and the scenario-based ambiguity set is very suitable, as evidenced by two factors: First, the empirical scenario set can be easily obtained using the probability estimation results. Second, the errors of the probability estimations can be used to determine the parameters $\eta^{\max}$ and $\eta^{\min}$ in (13). More specifically, $\eta^{\max}$ and $\eta^{\min}$ can be set as the maximum and minimum deviations between the target and actual cumulative probabilities of the predicted quantiles, as:

$$\eta^{\max} = \max_{1 \leq j \leq \mathcal{J}} \left\{ q_j^* - q_j^0, 0 \right\} \quad (14)$$

$$\eta^{\min} = \min_{1 \leq j \leq \mathcal{J}} \left\{ q_j^* - q_j^0, 0 \right\} \quad (15)$$

where $j$ and $\mathcal{J}$ are the index and the number of the predicted quantiles, $q_j^*$ and $q_j^0$ are the target and actual cumulative probabilities of the $j^{th}$ predicted quantile.

### C. DRO-Based AGC Signal Optimization Model

Based on the MPC technology and time-domain transform function (12), the AGC signal optimization model considering the uncertainties of the FRM key parameters can be formulated. In this study, it is assumed that the forecasts of the power fluctuations in the optimization horizon (four AGC periods) are available, which is similar to the existing researches on the model predictive-based AGC [6]–[10].

$$\textbf{P2:} \quad \min_{\boldsymbol{u}} \left[ \sum_{z=0}^{Z-1} C^{\text{R}} \cdot \Delta P_z^{\text{R}^2} + \max_{\mathcal{P} \in \widehat{\mathcal{P}}} \mathbb{E}^{\mathcal{P}} \left( \sum_{z=1}^{Z} C^{\text{f}} \cdot \Delta f_z^2 \right) \right],$$

$$\text{s.t.:} \quad \boldsymbol{x}_{z+1} = \boldsymbol{\mathcal{A}}^{\text{sec}} \boldsymbol{x}_z + \boldsymbol{\mathcal{B}}^{\text{sec}} \boldsymbol{u}_z, \forall 0 \leq z \leq Z-1, \quad (16)$$

$$\min_{\mathcal{P} \in \widehat{\mathcal{P}}} \mathbb{P}^{\mathcal{P}} \left( \underline{\Delta f} \leq \Delta f_z \leq \overline{\Delta f} \right) \geq \beta, \forall 1 \leq z \leq Z, \quad (17)$$

$$\underline{\Delta P^{\text{R}}} \leq \Delta P_z^{\text{R}} \leq \overline{\Delta P^{\text{R}}}, \forall 0 \leq z \leq Z-1. \quad (18)$$

where $z$ and $Z$ denote the index and number of the AGC periods in the optimization horizon, $C^{\text{R}}$ is the control cost coefficient, $C^{\text{f}}$ is the penalty coefficient for frequency deviation, $\mathcal{P}$ denotes the probability distribution, and $\mathbb{P}$ denotes the probability.

The optimization problem **P2** is a min-max bilevel optimization problem. The inner maximization problem is reformulated as follows:

$$\textbf{P3:} \quad \max_{\boldsymbol{\omega}} \sum_{j=1}^{\mathcal{J}} \omega_j \cdot \left( \sum_{z=1}^{Z} C^{\text{f}} \cdot \Delta f_{j,z}^2 \right)$$

$$\text{s.t.:} \quad \omega_j - \omega_j^0 \geq \eta^{\min}, \forall j, \qquad : \underline{\mu}_j \quad (19)$$

$$\omega_j - \omega_j^0 \leq \eta^{\max}, \forall j, \qquad : \overline{\mu}_j \quad (20)$$

$$\sum_{j=1}^{\mathcal{J}} \omega_j = 1, \qquad : \upsilon, \quad (21)$$

$$\omega_j \geq 0, \forall j. \quad (22)$$

where $\underline{\mu}_j$, $\overline{\mu}_j$, and $\upsilon$ are the dual variables corresponding to different constraints. To solve this problem, the inner maximization problem should be dualized first, as:

$$\textbf{P4:} \quad \min \left( \sum_{j=1}^{\mathcal{J}} \left[ (\omega_j^0 + \eta^{\min}) \cdot \underline{\mu}_j + (\omega_j^0 + \eta^{\max}) \cdot \overline{\mu}_j \right] + \upsilon \right)$$

$$\text{s.t.:} \quad \underline{\mu}_j + \overline{\mu}_j + \upsilon \geq \sum_{z=1}^{Z} C^{\text{f}} \cdot \Delta f_{j,z}^2, \forall j, \quad (23)$$

$$\underline{\mu}_j \leq 0, \overline{\mu}_j \geq 0, \forall j. \quad (24)$$

The constraint of the system frequency in (17) is a joint chance constraint, and it can be split into two separated constraints according to Bernoulli's inequality [27], as

$$\begin{cases} \min_{\mathcal{P} \in \widehat{\mathcal{P}}} \mathbb{P}^{\mathcal{P}} \left( \underline{\Delta f} \leq \Delta f_z \right) \geq 1 - \frac{\beta}{2} \\ \min_{\mathcal{P} \in \widehat{\mathcal{P}}} \mathbb{P}^{\mathcal{P}} \left( \Delta f_z \leq \overline{\Delta f} \right) \geq 1 - \frac{\beta}{2} \end{cases} \quad (25)$$

The transformation described above is conservative, which means that constraint (25) is a sufficient but unnecessary condition of constraint (17).

Conditional Value at Risk (CVaR) is a risk measure quantifies the expected loss over the portion of distribution that lies beyond the confidence level [28], [29]. After the CVaR approximation, the chance constraint (25) is expressed as follows:

$$\max_{\mathcal{P} \in \widehat{\mathcal{P}}} \text{CVaR}_{1-\beta/2} \left( \underline{\Delta f} \leq \Delta f_z \right)$$

$$= \begin{cases} \max_{\boldsymbol{\omega}_z} \left[ \dfrac{\sum\limits_{j=1}^{\mathcal{J}} \omega_{j,z} \cdot (\underline{\Delta f} - \Delta f_{j,z} - \underline{\delta}_z)^+}{1 - \beta/2} + \underline{\delta}_z \right] \leq 0 \\ \text{s.t.:} \ \omega_{j,z} \text{ satisfied constraints } (19) - (22), \end{cases} \quad (26)$$

and

$$\max_{\mathcal{P} \in \widehat{\mathcal{P}}} \text{CVaR}_{1-\beta/2} \left( \Delta f_z \leq \overline{\Delta f} \right)$$

$$= \begin{cases} \max_{\boldsymbol{\omega}_z} \left[ \dfrac{\sum\limits_{j=1}^{\mathcal{J}} \omega_{j,z} \cdot (\Delta f_{j,z} - \overline{\Delta f} - \bar{\delta}_z)^+}{1 - \beta/2} + \bar{\delta}_z \right] \leq 0 \\ \text{s.t.:} \ \omega_{j,z} \text{ satisfied constraints } (19) - (22). \end{cases} \quad (27)$$

By utilizing the dualization method, the above chance constraints can be finally transformed into the following form (take constraint (26) as an example):

$$\begin{cases} \sum_{j=1}^{\mathcal{J}} \left[ (\omega_j^0 + \eta^{\min}) \cdot \underline{\hbar}_{j,z} + (\omega_j^0 + \eta^{\max}) \cdot \overline{\hbar}_{j,z} \right] + \\ \qquad\qquad \underline{\lambda}_z + \underline{\delta}_z \leq 0, \forall 1 \leq z, \\ \underline{\hbar}_{j,z} + \overline{\hbar}_{j,z} + \underline{\lambda}_z \geq \underline{\vartheta}_{j,z}, \qquad \forall j, 1 \leq z, \\ \underline{\vartheta}_{j,z} \geq \dfrac{(\underline{\Delta f} - \Delta f_{j,z} - \underline{\delta}_z)}{1 - \beta/2}, \forall j, 1 \leq z, \\ \underline{\hbar}_{j,z} \leq 0, \overline{\hbar}_{j,z} \geq 0, \underline{\vartheta}_{j,z} \geq 0, \forall j, 1 \leq z. \end{cases} \quad (28)$$

where $\underline{\hbar}_{j,z}$, $\overline{\hbar}_{j,z}$, and $\underline{\lambda}_{j,z}$ are the dual variables corresponding to the constraints for $\omega_{j,z}$ in (26), $\underline{\vartheta}_{j,z}$ is an auxiliary variable. The derivation process is given in Appendix B.

After the above transformations, the original optimization problem **P1** is reformulated as follows:

**P5:** $\min \left( \sum_{j=1}^{\mathcal{J}} \left[ (\omega_j^0 + \eta^{\min}) \cdot \underline{\mu}_j + (\omega_j^0 + \eta^{\max}) \cdot \overline{\mu}_j \right] + \sum_{z=0}^{Z-1} C^{\text{R}} \cdot \Delta P_z^{\text{R}^2} + v \right)$

s.t.: $x_{j,z+1} = \boldsymbol{\mathcal{A}}_j^{\text{sec}} x_{j,z} + \boldsymbol{\mathcal{B}}_j^{\text{sec}} u_z, \forall 0 \leq z \leq Z-1, j,$ (29)

$\underline{\mu}_j + \overline{\mu}_j + v \geq \sum_{z=1}^{Z} C^{\text{f}} \cdot \Delta f_{j,z}^2, \forall j,$ (30)

$\sum_{j=1}^{\mathcal{J}} \left[ (\omega_j^0 + \eta^{\min}) \cdot \underline{\phi}_{j,z} + (\omega_j^0 + \eta^{\max}) \cdot \overline{\phi}_{j,z} \right]$
$\qquad\qquad + \psi_z + \overline{\delta}_z \leq 0, \forall j, 1 \leq z,$ (31)

$\underline{\phi}_{j,z} + \overline{\phi}_{j,z} + \psi_z \geq \overline{\vartheta}_{j,z}, \forall j, 1 \leq z,$ (32)

$\overline{\vartheta}_{j,z} \geq \dfrac{(\Delta f_{j,z} - \overline{\Delta f} - \overline{\delta}_z)}{1 - \beta/2}, \forall j, 1 \leq z,$ (33)

$\overline{\vartheta}_{j,z} \geq 0, \underline{\phi}_{j,z} \leq 0, \overline{\phi}_{j,z} \geq 0, \underline{\mu}_j \leq 0, \overline{\mu}_j \geq 0, \forall j, 1 \leq z,$ (34)

$\underline{\Delta P^{\text{R}}} \leq \Delta P_z^{\text{R}} \leq \overline{\Delta P^{\text{R}}}, \forall 0 \leq z \leq Z-1,$ (35)

$\text{constraints} \quad (28).$ (36)

Constraints (31)-(34) are transformed from chance constraint (27), where $\underline{\phi}_{j,z}$, $\overline{\phi}_{j,z}$, and $\psi_z$ are the dual variables of the constraints for $\omega_{j,z}$. $\vartheta_{j,z}$ is an auxiliary variable.

**P5** is a quadratically constrained quadratic programming problem, which can be efficiently solved by the existing optimization algorithms.

## V. Case Study

### A. Case Settings

The modified IEEE 118-bus system is used as the test system [30], as shown in Fig. 4. Six wind farms with installed capacities of 400, 300, 300, 330, 400, and 300 MW, are connected to buses 1, 44, 68, 17, 76, and 89, respectively. Four photovoltaic power stations with installed capacities of 400, 400, 330, and 300 MW are linked to buses 18, 24, 32, and 38, respectively. Moreover, each renewable power station is outfitted with a BESS with a power capacity equal to 10 % of the corresponding renewable generation.

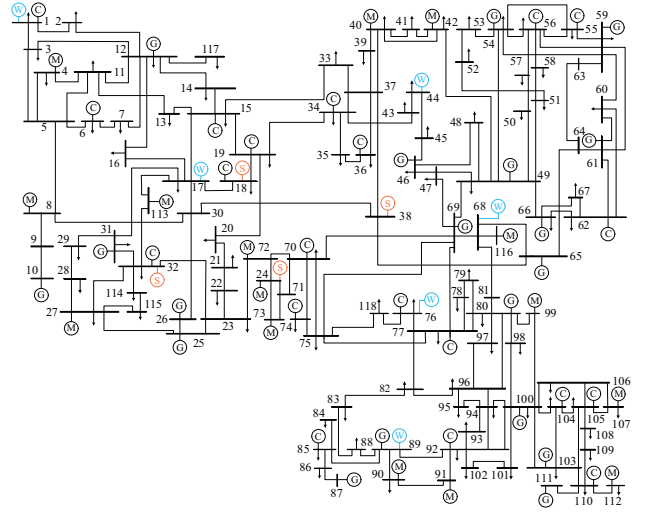

Fig. 4. Diagram of the IEEE 118-bus test system.

The TG parameters in the FRM are taken from reference [31], and the BESS converter parameters are given in Table I. Renewable generations do not participate the frequency regulation, and the AGC signal is allocated among the TGs and BESSs based on their participation factors.

TABLE I
CONVERTER PARAMETERS OF THE BESSs

| Bus | Converter Time Constant | Droop Factor |
| --- | --- | --- |
| 1,44,68,17,76,89 | 0.01 s | 0.067 |
| 18,24,32,38 | 0.015 s | 0.08 |

The historical data of load power, wind and solar generations used in this study are acquired from an actual power system in North China with a time span from January 2016 to March 2018. All the historical data are scaled down for the simulation. Because the sampling period is too long (one minute) for the frequency control simulation, linear interpolation is used to convert the historical data into a smaller time resolution.

Based on the unit commitment model proposed in section 4.2.3 of reference [32], the on-off schedule of TGs is simulated by utilizing the scaled load power and renewable generation data as the input. The on-off statuses of TGs are used in the tests of the proposed FRM key parameter identification method and AGC signal optimization method.

### B. Identification Results of FRM Key Parameters

Because the actual system inertia and load damping factor are not available in practice, the artificially generated data of these parameters are used. The system inertia is contributed by the online TGs and load. The former's inertia is computed using the TGs' installed capacities and inertia coefficients, and the latter's inertia is calculated by utilizing the load power and inertial coefficient, which is assumed to obey the normal distribution $N(1.79, 0.31)$ according to [33]. The load damping factor is sampled from the normal distribution $N(1\%, 0.3\%)$ [34], [35].

In the case study, the proposed parameter identification method is implemented in five steps. First, in historical data, the time periods with a netload power shift of more than 2% within one minute are chosen as the ramp events. Second, the system inertia and load damping factor data are randomly generated using the methods mentioned in the preceding paragraph. Third, the power disturbance data of power ramp events, generated FRM key parameters, and the on-off statuses of TGs are inputted into the FRM to get the frequency response simulation results. Fourth, the frequency results obtained from the simulation are used for the offline parameter identifications. Finally, based on the offline identification results, the online probability estimate model of the FRM key parameters is trained.

After processing the historical netload data, 4241 ramp events are found. The online probability estimation model is trained by using 3000 ramp events as the training set, 1000 ramp events as the validation set, and 241 ramp events as the test set.

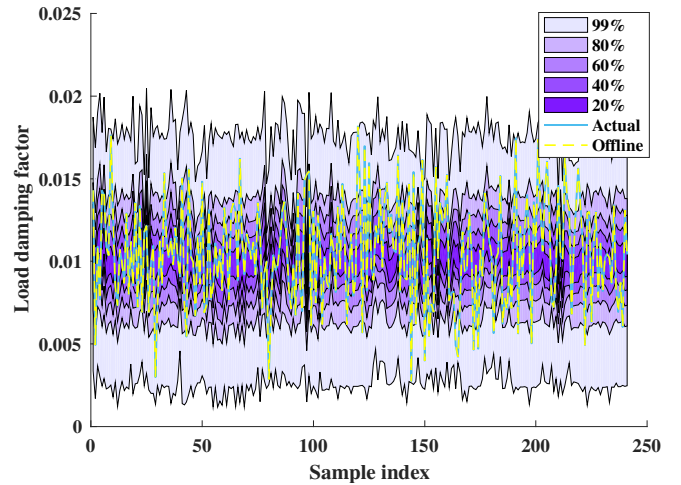

Fig. 6. Offline and online identification results of the load damping factor.

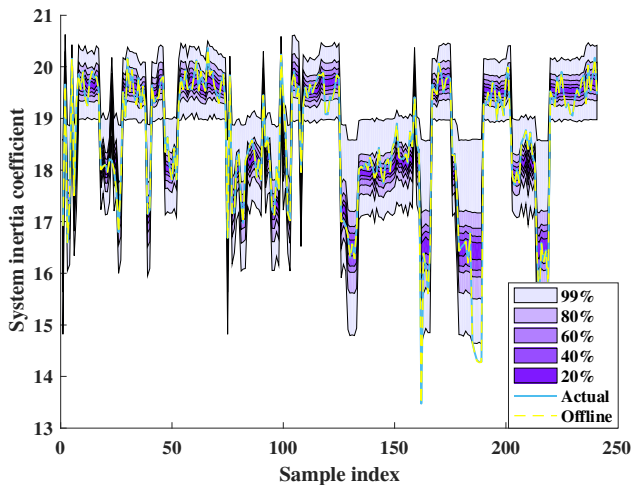

Fig. 5. Offline and online identification results of the system inertia coefficient.

Figs. 5 and 6 show the offline and online identification results of the system inertia coefficient and load damping factor on the test set. The offline identification results are close to their actual values. The root mean square errors of the system inertia coefficient and load damping factor are 0.01 and $2.5 \times 10^{-4}$ respectively, indicating that the offline identification method has high accuracy.

The colored zones represent the results of online probability estimations. It can be intuitively found from the figures that the probability estimation results are comparatively accurate. To quantitatively analyze the errors of online probability estimations, the differences between the target and actual cumulative probabilities of the predicted quantiles are calculated. Table II shows the comparison results of the quantiles with target cumulative probabilities ranging from 10% to 90% at a 20% interval. The maximum absolute deviations for the system inertia coefficient and load damping factor are 3.94% and 1.62%, respectively, implying that the online probability estimation method has good accuracy.

TABLE II
PROBABILITY ESTIMATION DEVIATIONS OF THE FRM KEY PARAMETERS

| Target cumulative probabilities | Deviation of system inertia coefficient | Deviation of load damping factor |
|---|---|---|
| 10% | -1.62% | -1.70% |
| 30% | -0.29% | -3.03% |
| 50% | 0.21% | -3.94% |
| 70% | -0.71% | -3.61% |
| 90% | -0.37% | -2.45% |

### C. Comparisons of Different Methods

In this subsection, the proposed AGC method is compared with the traditional PI-based AGC method in the condition of varying FRM key parameters.

For the traditional PI-based AGC method, the variations in system inertia and load damping factor are not considered. The proportional and integral gains $K_p$ and $K_i$ are turned through the exhaustive search to minimize the objective function in **P2**. The values of the system inertia and load damping factor used in the FRM for turning the controller gains are their average values in the training and validation sets. The netload disturbance used in the simulation is depicted in Fig. 7.

Three typical cases are further selected from the test set according to the system inertia and load damping factor, and they are utilized to test each AGC method. $C^R$ and $C^f$ in **P2** are taken as 30 and 15000 in the case study.

Fig. 8(a)-(c) plot the system frequency and regulation signal in the simulation period. The simulation results of the first case are shown in Fig. 8(a). In this circumstance, the system inertia coefficient and load damping factor are lower than their average values. As a result, the PI-based control technique generates insufficient regulation signals. Furthermore, its frequency deviations exceed the limit, when the power disturbances are comparatively large (from 75s to 89s).

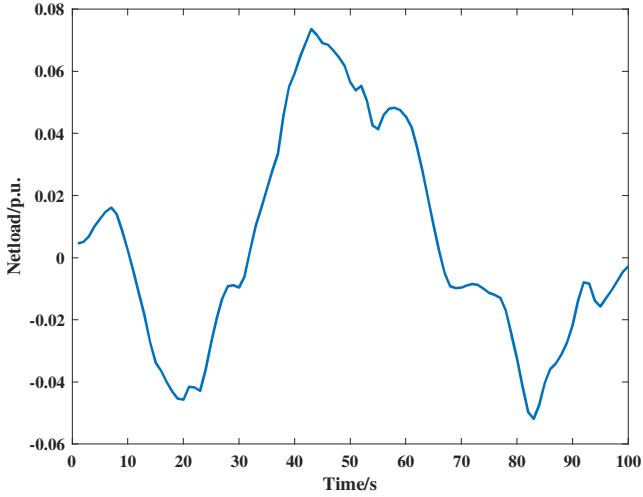

Fig. 7. Netload disturbance used in the simulation.

In the second case, the system inertia coefficient and load damping factor are both close to their average values, respectively. As shown in 8(b), the regulation signals and the frequency deviations of the PI-based control method are similar to those of the proposed control method, demonstrating that the control performance of the proposed method is equivalent to that of a well-turned PI-based controller with the FRM key parameters known.

Fig. 8(c) gives the simulation results when the system inertia coefficient and load damping factor are greater than their average values. In this case, the AGC signals produced by the PI-based based method are larger than those produced by the proposed method. Nonetheless, the improvement in the frequency control performance is minor for the former, and the frequency deviations of the later are also within the limit.

TABLE III
COMPARISONS ON THE PROPOSED AND PI-BASED CONTROL METHODS

| Control method | Regulation signal | Frequency deviation | Objective value | Frequency out-of-limit |
| --- | --- | --- | --- | --- |
| Proposed | 0.0102 p.u. | 0.0290 Hz | 0.0042 | 0% |
| PI-based | 0.0103 p.u. | 0.0312 Hz | 0.0063 | 2% |

Table III shows the simulation results for both control methods, where the regulation signal, frequency deviation, and objective value are the mean absolute value. The frequency out-of-limit percentage is the average proportion of frequency out-of-limit over three test cases. The following are some of the conclusions drawn from the Table III:

1) The average regulation signal of the proposed method is *slightly smaller* than that of the PI-based control method.
2) The proposed method has *lower* average frequency deviation than the PI-based control method.
3) The mean objective value of the proposed method is *notably smaller* than that of the PI-based control method.
4) The frequency out-of-limit percentage is zero for the proposed method, while this percentage is 2% for the PI-based control method.

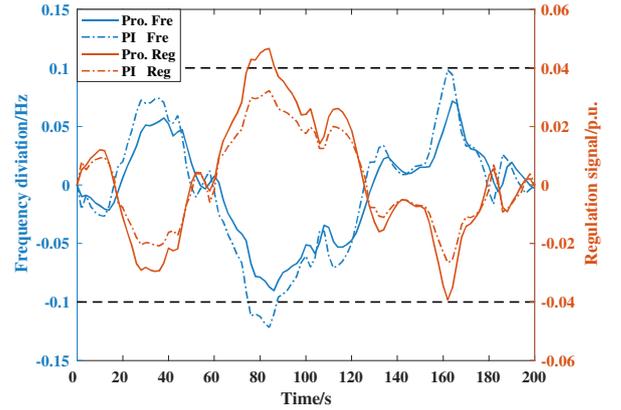

(a) System inertia coefficient = 13.48, load damping factor = 0.0041.

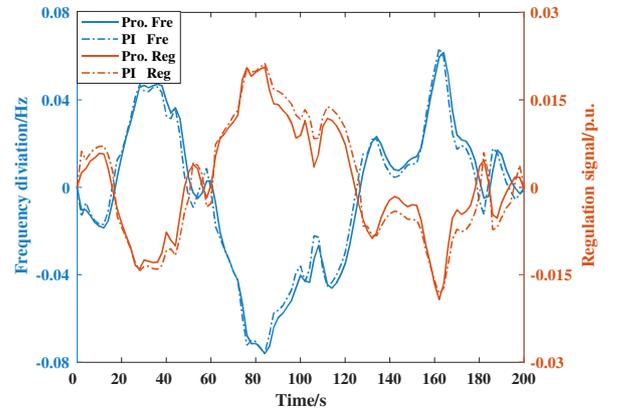

(b) System inertia coefficient = 17.74, load damping factor = 0.0105.

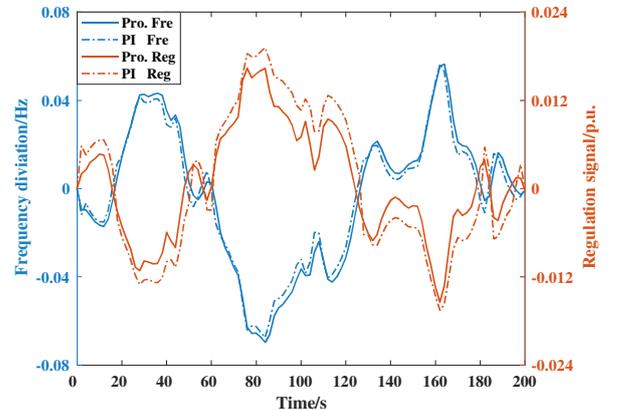

(c) System inertia coefficient = 20.00, load damping factor = 0.0174.

Fig. 8. Simulation results under different system inertia coefficients and load damping factors.

These comparison results show that the proposed method achieves better frequency control performance at a lower control cost.

### D. Computation Efficiency of the Proposed Method

One AGC period typically lasts 2-8 seconds, hence the computation efficiency of an AGC method is very important.

In this paper, the experiment is carried on a desktop PC with an Intel(R) Core(TM) i7-7700 CPU 3.6 GHz and 8 GB of memory. The average computation time for the proposed control method is **0.21** seconds, which is one order smaller than the AGC period. Furthermore, the computation efficiency can be further improved by using more powerful computing resources. Thus, the computational time for the proposed AGC method is acceptable for real implementation.

## VI. Conclusion

This paper proposes a model predictive-based AGC method considering the uncertainties of the FRM key parameters. A systematic parameter identification method is proposed to estimate the probability distributions of the FRM key parameters. On this basis, the DRO technique is utilized to develop the AGC signal optimization model. The simulation results reveal that the proposed control method outperforms the widely used PI-based control method in terms of both control performance and cost. In future work, studies will be carried out to extant the proposed method to the interconnected power systems with multiple control areas.

## Appendix A

The matrices $\mathcal{A}$ and $\mathcal{B}$ in the sate space equation (1) are given as follows:

$$\mathcal{A} = \begin{bmatrix} \mathcal{A}^{\mathrm{s}} & \mathcal{A}^{\mathrm{g2s}}_1 & \cdots & \mathcal{A}^{\mathrm{g2s}}_{1+N^{\mathrm{rh}}} & \cdots & \mathcal{A}^{\mathrm{e2s}}_1 & \cdots & \mathcal{A}^{\mathrm{e2s}}_{N^{\mathrm{es}}} \\ \mathcal{A}^{\mathrm{s2g}}_1 & \mathcal{A}^{\mathrm{g}}_1 & 0 & 0 & 0 & 0 & 0 & 0 \\ \vdots & \vdots & \ddots & \vdots & \vdots & \vdots & \vdots & \vdots \\ \mathcal{A}^{\mathrm{s2g}}_{1+N^{\mathrm{rh}}} & 0 & 0 & \mathcal{A}^{\mathrm{g}}_{1+N^{\mathrm{rh}}} & 0 & 0 & 0 & 0 \\ \vdots & 0 & 0 & 0 & \ddots & 0 & 0 & 0 \\ \mathcal{A}^{\mathrm{s2e}}_1 & 0 & 0 & 0 & 0 & \mathcal{A}^{\mathrm{e}}_1 & 0 & 0 \\ \vdots & 0 & 0 & 0 & 0 & 0 & \ddots & 0 \\ \mathcal{A}^{\mathrm{s2e}}_{N^{\mathrm{es}}} & 0 & 0 & 0 & 0 & 0 & 0 & \mathcal{A}^{\mathrm{e}}_{N^{\mathrm{es}}} \end{bmatrix}$$

$$\mathcal{A}^{\mathrm{s}} = -\frac{D}{2H}, \qquad \mathcal{A}^{\mathrm{g2s}}_1 = \begin{bmatrix} \frac{1}{2H} & 0 & 0 \end{bmatrix},$$

$$\mathcal{A}^{\mathrm{e2s}}_1 = \frac{1}{2H}, \qquad \mathcal{A}^{\mathrm{g2s}}_{1+N^{\mathrm{rh}}} = \begin{bmatrix} \frac{1}{2H} & 0 \end{bmatrix},$$

$$\mathcal{A}^{\mathrm{s2e}}_1 = -\frac{1}{R^{\mathrm{V}}T^{\mathrm{V}}}, \qquad \mathcal{A}^{\mathrm{s2g}}_{1+N^{\mathrm{rh}}} = \begin{bmatrix} 0 & -\frac{1}{T^{\mathrm{G}} \cdot R_{1+N^{\mathrm{rh}}}} \end{bmatrix}^{\top},$$

$$\mathcal{A}^{\mathrm{s2g}}_1 = \begin{bmatrix} 0 & 0 & -\frac{1}{T^{\mathrm{G}} \cdot R_1} \end{bmatrix}^{\top},$$

$$\mathcal{A}^{\mathrm{g}}_1 = \begin{bmatrix} -\frac{1}{T^{\mathrm{R}}} & \frac{T^{\mathrm{C}} - F^{\mathrm{H}}T^{\mathrm{R}}}{T^{\mathrm{R}}T^{\mathrm{C}}} & \frac{F^{\mathrm{H}}}{T^{\mathrm{C}}} \\ 0 & -\frac{1}{T^{\mathrm{C}}} & \frac{1}{T^{\mathrm{C}}} \\ 0 & 0 & -\frac{1}{T^{\mathrm{G}}} \end{bmatrix},$$

$$\mathcal{A}^{\mathrm{g}}_{1+N^{\mathrm{rh}}} = \begin{bmatrix} -\frac{1}{T^{\mathrm{C}}} & \frac{1}{T^{\mathrm{C}}} \\ 0 & -\frac{1}{T^{\mathrm{G}}} \end{bmatrix}, \qquad \mathcal{A}^{\mathrm{e}}_1 = -\frac{1}{R^{\mathrm{V}}T^{\mathrm{V}}}.$$

$$\mathcal{B} = \begin{bmatrix} 0 & \frac{\alpha^{\mathrm{G}}_1}{T^{\mathrm{G}}_1} & \cdots & \frac{\alpha^{\mathrm{G}}_{1+N^{\mathrm{rh}}}}{T^{\mathrm{G}}_{1+N^{\mathrm{rh}}}} & \cdots & \frac{\alpha^{\mathrm{V}}_1}{T^{\mathrm{V}}_1} & \cdots & \frac{\alpha^{\mathrm{V}}_{N^{\mathrm{es}}}}{T^{\mathrm{V}}_{N^{\mathrm{es}}}} \\ -\frac{1}{2H} & 0 & \cdots & 0 & \cdots & 0 & \cdots & 0 \end{bmatrix}^{\top}$$

The matrices $\mathcal{A}^{\mathrm{pri}}$, $\mathcal{B}^{\mathrm{pri}}$, and $\boldsymbol{u}_k$ in the transform function (7) are given as follows:

$$\mathcal{A}^{\mathrm{pri}} = \mathrm{e}^{\mathcal{A}\Gamma}$$

$$\mathcal{B}^{\mathrm{pri}} = \begin{bmatrix} \int_0^{\Gamma} \mathrm{e}^{\mathcal{A}(\Gamma-\tau)} \mathcal{B}[2] \mathrm{d}\tau & \dfrac{\int_0^{\Gamma} \mathrm{e}^{\mathcal{A}(\Gamma-\tau)} \mathcal{B}[2] \tau \mathrm{d}\tau}{\Gamma} \end{bmatrix}$$

$$\boldsymbol{u}_k = \begin{bmatrix} \Delta P^{\mathrm{NL}}_k & \Delta P^{\mathrm{NL}}_{k+1} - \Delta P^{\mathrm{NL}}_k \end{bmatrix}^{\top}$$

The matrices $\mathcal{A}^{\mathrm{sec}}$, $\mathcal{B}^{\mathrm{sec}}$, and $\boldsymbol{u}_z$ in the transform function (12) are given as follows:

$$\mathcal{A}^{\mathrm{sec}} = \mathrm{e}^{\mathcal{A}\Upsilon}$$

$$\mathcal{B}^{\mathrm{sec}} = \begin{bmatrix} \int_0^{\Upsilon} \mathrm{e}^{\mathcal{A}(\Upsilon-\tau)} \mathcal{B}[1] \mathrm{d}\tau & \int_0^{\Upsilon} \mathrm{e}^{\mathcal{A}(\Upsilon-\tau)} \mathcal{B}[2] \mathrm{d}\tau \\ & \dfrac{\int_0^{\Upsilon} \mathrm{e}^{\mathcal{A}(\Upsilon-\tau)} \mathcal{B}[2] \tau \mathrm{d}\tau}{\Upsilon} \end{bmatrix}$$

$$\boldsymbol{u}_z = \begin{bmatrix} \Delta P^{\mathrm{R}}_z & \Delta P^{\mathrm{NL}}_z & \Delta P^{\mathrm{NL}}_{z+1} - \Delta P^{\mathrm{NL}}_z \end{bmatrix}^{\top}$$

## Appendix B

The chance constraint (26) can be reformulated as follows:

$$\textbf{P6:} \quad \max_{\boldsymbol{\omega}_z} \frac{\sum\limits_{j=1}^{\mathcal{J}} \omega_{j,z} \cdot \left( \Delta f - \Delta f_{j,z} - \underline{\delta}_z \right)^+}{1 - \beta/2}$$

$$\text{s.t.:} \quad \omega_{j,z} - \omega^0_{j,z} \geq \eta^{\min}, \forall j, 1 \leq z, : \underline{\hbar}_{j,z} \qquad (37)$$

$$\omega_{j,z} - \omega^0_{j,z} \leq \eta^{\max}, \forall j, 1 \leq z, : \overline{\hbar}_{j,z} \qquad (38)$$

$$\sum_{j=1}^{\mathcal{J}} \omega_{j,z} = 1, \forall j, 1 \leq z, : \lambda_z, \qquad (39)$$

$$\omega_{j,z} \geq 0, \forall j, 1 \leq z. \qquad (40)$$

where $\underline{\delta}_z$ is the auxiliary variables in the CVaR approximation, $\underline{\hbar}_{j,z}$, $\overline{\hbar}_{j,z}$, and $\lambda_z$ are dual variables. The dual form of the above maximization problem is given as below:

$$\textbf{P7:} \quad \min \left( \sum_{j=1}^{\mathcal{J}} \left[ (\omega^0_j + \eta^{\min}) \cdot \underline{\hbar}_{j,z} + (\omega^0_j + \eta^{\max}) \cdot \overline{\hbar}_{j,z} \right] + \lambda_z \right)$$

$$\text{s.t.:} \quad \underline{\hbar}_{j,z} + \overline{\hbar}_{j,z} + \lambda_z \geq \frac{(\Delta f - \Delta f_{j,z} - \underline{\delta}_z)^+}{1 - \beta/2}, \forall j, 1 \leq z, \qquad (41)$$

$$\underline{\hbar}_{j,z} \leq 0, \overline{\hbar}_{j,z} \geq 0, \forall j, 1 \leq z. \qquad (42)$$

where the expression $\left(\underline{\Delta f} - \Delta f_{j,z} - \underline{\delta}_z\right)^+$ can be transformed into

$$\begin{cases} \underline{h}_{j,z} + \overline{h}_{j,z} + \check{\lambda}_z \geq \underline{\vartheta}_{j,z}, \forall j, 1 \leq z, \\ \underline{\vartheta}_{j,z} \geq \dfrac{\left(\underline{\Delta f} - \Delta f_{j,z} - \underline{\delta}_z\right)}{1 - \beta/2}, \forall j, 1 \leq z, \\ \underline{\vartheta}_{j,z} \geq 0, \forall j, 1 \leq z, \end{cases} \quad (43)$$

This transformation is equivalent because the factors of $\underline{h}_{j,z}$, $\overline{h}_{j,z}$, and $\check{\lambda}_z$ in the objective function are all positive. Then, the chance constraint (26) can be expressed as follows:

$$\begin{cases} \min\left(\sum\limits_{j=1}^{\mathcal{J}}\left[(\omega_j^0 + \eta^{\min}) \cdot \underline{h}_{j,z} + (\omega_j^0 + \eta^{\max}) \cdot \overline{h}_{j,z}\right] \right. \\ \left. + \check{\lambda}_z + \underline{\delta}_z \right) \leq 0, \forall j, 1 \leq z \\ \underline{h}_{j,z} + \overline{h}_{j,z} + \check{\lambda}_z \geq \underline{\vartheta}_{j,z} \forall j, 1 \leq z, \\ \underline{\vartheta}_{j,z} \geq \dfrac{\left(\underline{\Delta f} - \Delta f_{j,z} - \underline{\delta}_z\right)}{1 - \beta/2}, \forall j, 1 \leq z, \\ \underline{\vartheta}_{j,z} \geq 0, \forall j, 1 \leq z, \end{cases}$$
(44)

where the operator min can be omitted directly since if the value of an expression can be smaller than 0, the minimum of the expression is certainly smaller than 0.